\documentclass[aps,prl,twocolumn,showpacs,notitlepage,floatfix,longbibliography,superscriptaddress]{revtex4-1}

\usepackage{amsmath,amssymb}
\usepackage{graphicx}
\usepackage{xcolor}
\usepackage{hyperref}
\usepackage{physics}
\usepackage{dsfont}
\usepackage{footmisc}
\usepackage{cleveref}
\hypersetup{colorlinks=true,citecolor={blue},linkcolor={blue},urlcolor={blue}}

\newcommand{\pksadd}{Max Planck Institute for the Physics of Complex Systems, Dresden D-01187, Germany}
\newcommand{\pcsadd}{Center for Theoretical Physics of Complex Systems, Institute for Basic Science (IBS), Daejeon 34126, Republic of Korea}
\newcommand{\ustadd}{Basic Science Program, Korea University of Science and Technology (UST), Daejeon, Korea, 34113}

\newcommand{\sect}[1]{\textsl{#1} ---}

\newcommand{\mh}{\hat{\ensuremath{\mathcal{H}}}}
\newcommand{\mhsp}{\hat{\ensuremath{\mathcal{H}}}_\text{sp}}
\newcommand{\mhi}{\hat{\ensuremath{\mathcal{H}}}_\text{int}}

\newcommand{\ml}{\mathbf{l}}

\newcommand{\ha}{\hat{a}}
\newcommand{\hb}{\hat{b}}
\newcommand{\hc}{\hat{c}}

\newcommand{\hn}{\hat{n}}

\newcommand{\ho}{\hat{I}}

\begin{document}

\title{Heat percolation in many-body flatband localizing systems
%Percolation transitions in interacting many-body flatband systems
% {\small Buzzwords: Many-body system; Disorder-free localization;
% Many-body localization;
% Flat bands; 
% Percolations; 
% Interaction-induced transport; 
% Ising model; 
% Transport; 
% Hilbert space fragmentation
% }
}

\author{I.~Vakulchyk}
\email[Corresponding author\\]{igrvak@gmail.com}
\affiliation{\pcsadd}
\affiliation{\ustadd}

\author{C.~Danieli}
\email[Corresponding author\\]{cdanieli@pks.mpg.de}
\affiliation{\pksadd}

\author{Alexei~Andreanov}
\email[Corresponding author\\]{aalexei@ibs.re.kr}
\affiliation{\pcsadd}
\affiliation{\ustadd}

\author{S.~Flach}
\email[Corresponding author\\]{sergejflach@googlemail.com}
\affiliation{\pcsadd}
\affiliation{\ustadd}

\begin{abstract}
Translationally invariant finetuned single-particle lattice Hamiltonians host flat bands only.  Suitable short-range many-body interactions result in complete suppression of particle transport due to local constraints and Many-Body Flatband Localization.  Heat can still flow between spatially locked charges.
We show that heat transport is forbidden in dimension one. In higher dimensions heat transport can be unlocked by tuning filling fractions across a
percolation transition
for suitable lattice geometries.
%We show that the possibility of heat transport is governed by percolation mechanism in real space.
%In one-dimension, conducting channels are either completely absent or their number is exponentially suppressed with the system size.
%In higher dimensions, the number of conducting channels can undergo a percolation transition depending on the geometry and upon tuning the filling fraction, potentially triggering heat transport.
Transport in percolation clusters is additionally affected by effective bulk disorder and edge scattering induced by the local constraints, which
work in favor of arresting the heat flow.
We discuss explicit examples in one and two dimensions. 
\end{abstract}

\date{\today}

\maketitle

\sect{Introduction} 
The breaking of ergodicity in interacting quantum many-body systems, known as many-body localization (MBL), is an important open problem and active topic of research.
%The celebrated many-body localization (MBL) is one of the most notable examples of the impressive set of developments achieved in the field of quantum many-body systems during the last decades. 
The first studies~\cite{fleishman1980interaction,altshuler1997quasi,jacquod1997emergence,gornyi2005interacting,basko2006metal} showed that MBL and the exponential suppression of any transport arise in one-dimensional interacting systems from the interplay of random fields and interactions. 
While the original MBL framework has been widely developed both theoretically and experimentally~\cite{abanin2017recent,abanin2019colloquium}, the need for random fields as a key element for MBL has been later relaxed as MBL features have been found in a variety of systems without disorder~\cite{schiulaz2015dynamics,vanhorssen2015dynamics,pino2016nonergodic,hickey2016signatures,mondaini2017many,schulz2019stark} -- particularly, in interacting systems featuring an extensive number of local constrains~\cite{smith2017disorder,smith2017absence,smith2018dynamical,brenes2018many,hart2020logarithmic,karpov2020disorderfree}. 
Computationally, interacting quantum systems face an exponential divergence in complexity over the system size, which limits most of numerical efforts to 1D. This, and the lack of analytically treatable models and methods, renders the crucial quest for MBL in higher-dimensional networks particularly challenging. Indeed, while signatures of MBL in 2D systems have been reported~\cite{wahl2019signatures,choi2016exploring}, it has also been shown that in $D>1$ the MBL regime is possibly unstable and destabilised in the long time limit~\cite{roeck2017stability}.

In this Letter, we study the transport features of disorder-free many-body systems which are based on Hamiltonian networks completely lacking single-particle dispersion. 
In particular, we relate the existence of phase transitions between conducting and insulating regimes to the lattice profile and its dimensionality. 
In translationally invariant lattices, the lack of single-particle dispersion in \emph{e.g.} one Bloch band typically results in a macroscopically degenerate set of eigenstates, all spatially compact  (CLS)~\cite{derzhko2015strongly,leykam2018artificial,leykam2018perspective}. 
Hence, a complete lack of dispersion -- i.e. all Bloch bands are flat (ABF)~\cite{vidal1998aharonov,vidal2000interaction,doucot2002pairing} -- results in the absence of extended states and strict single-particle confinement in the network. 
Flatband networks are increasingly permeating the quantum many-body field. 
Indeed, single-particle CLS have been extended to many-body CLS~\cite{tovmasyan2018preformed,tilleke2020nearest,danieli2020quantum}, while quantum scars~\cite{hart2020compact,mcclarty2020disorder,kuno2020flat_qs} and MBL-like dynamics~\cite{daumann2020manybody,khare2020localized,danieli2021manybody} have been reported very recently in flatband lattices. 
But more importantly, it has been shown that fine-tuning of the interaction in ABF networks induces an extensive set of local constraints which completely suppress charge transport in any spatial dimension -- a phenomenon called many-body flatband localization (MBFBL)~\cite{danieli2020many,kuno2020flat,orito2020exact}.
We focus on MBFBL networks, and by recasting them into a site-percolation problem we show that the structure of the Hilbert space undergoes a percolation transition upon tuning the filling fraction. 
Percolation approaches have been formerly employed to study MBL phenomena -- \emph{e.g.} Refs.~\cite{roy2019percolation,roy2019exact,prelov2021many}, and more recently~\cite{karpov2020disorderfree}. 
One of our main findings is that the critical value depends only on the lattice geometry and dimensionality, but not on particular realizations of the Hamiltonian terms -- thus indicating universality. 
More specifically, such transitions are absent and any transport is suppressed in all 1D networks, whereas in $D \geq 2$ the presence of the transition depends on the MBFBL geometry -- highlighting high-dimensional many-body systems where transport vanishes at any filling fraction. 
Additionally, we found that in the conducting phases predicted by percolation, the local constraints hurdle and potentially stop transport by generating effective bulk and edge disorders.

\sect{Setup} 
We study interacting many-body systems whose Hamiltonians consist of $D$-dimensional networks with $\nu$ strictly flat energy bands equipped with fine-tuned two-body interactions. Without loss of generality we consider spinless fermions, and following the scheme outlined in Ref.~\cite{danieli2020many} we consider a set of Hamiltonians written as
\begin{eqnarray}
    \footnotesize
    \hspace*{-5mm}  
    &&\mh  =  \mhsp + V\mhi \nonumber\\
     &=& \sum_{\bf l} 
    \hat{C}_{\bf l}^{\dagger T} H_0\hat{C}_{\bf l}
    +V \sum_{\langle{\bf l_1},{\bf l_2}\rangle}\sum_{a,b} 
    J_{a,b}^{{\bf l_1},{\bf l_2}}\hn_{{\bf l_1},a} \hn_{{\bf l_2},b},
\label{eq:2Dhamilt}
\end{eqnarray} 
where in $\mhsp$ the fermionic annihilation (creation) operators $\hc_{{\ml},a} (\hc_{{\ml},a}^{\dagger}$) have been grouped in $\nu$-dimensional vectors $\hat{C}_{\ml}(\hat{C}_{\ml}^\dagger)$, while $H_0$ is a Hermitian matrix, ${\ml}$ is a $D$-dimensional multi-index and $a,b$ label sites in unit cells. 
In Ref.~\onlinecite{danieli2020many}, this representation of $\mhsp$ has been named \emph{semi-detangled} since $\mhsp$ consists of decoupled unit cells of $\nu$ lattice sites each, whose profile is defined by $H_0$. 
The interaction $\mhi$ is set as products of particle-number operators $\hn_{{\ml},a} = \hc_{{\ml},a}^\dagger \hc_{{\ml},a}$. 
The multi-indexes $\langle{\ml_1},{\ml_2}\rangle$ account for neighboring unit cells reflecting the geometry of the interaction network, while the coefficients $J_{a,b}^{{\ml_1},{\ml_2}}$ define the interaction between cells ${\ml_1}$ and ${\ml_2}$. 

The addition of a fine-tuned density-density interaction $\mhi$ to the semi-detangled all-band-flat networks $\mhsp$ in Eq.~\eqref{eq:2Dhamilt} results in strict particle localization -- a phenomenon labeled Many-Body Flatband Localization (MBFBL). 
Indeed, the local operators $\ho_{\ml} = \sum_{a=1}^{\nu} \hn_{{\ml},a}$ that  count the number of particles within each unit cell, commute with $\mh$~\eqref{eq:2Dhamilt} and prohibit charge transport. The Hilbert space decomposes into dynamically independent subspaces characterized by the eigenvalues of $\{\ho_{\ml}\}$.
In general, the local conserved quantities $\ho_{\ml}$ do not forbid global heat transport, since particles confined to neighboring unit cells might still  exchange energy locally via the interaction $\mhi$~\footnote{Let us observe that if $H_0$ in Eq.~\eqref{eq:2Dhamilt} is diagonal, then every particle number operator $\hn_{{\ml},a}$ for any ${\ml}$ and $a$ is conserved -- hence fully suppressing heat transport as well}.
However the local heat exchange vanishes if at least one of the two cells coupled by $\mhi$ is either empty or full, i.e. $\ho_{\ml} = \{0, \nu\}$. 
Consequently unit cells of Eq.~\eqref{eq:2Dhamilt} split according to their occupation number into \emph{blocking} (empty/full cell, $\ho_{\ml}=\{0,\nu\}$) or \emph{conducting} (all the other cases, $\ho_{\ml}\neq \{0,\nu\}$)
%-- i.e., a cell is blocking if it is empty/full  or it is conducting otherwise $\ho_{\bf l}\neq \{0,\nu\}$. 
The impact of blocking cells on the global heat transport in Eq.~\eqref{eq:2Dhamilt} depends on the network geometry encoded by $\langle{\ml_1},{\ml_2}\rangle$, its spatial dimension $D$, the number of single particle flatbands $\nu$, and the filling fraction $\delta$. 

\sect{One-dimensional case}
In 1D MBFBL networks~\eqref{eq:2Dhamilt}, a single blocking unit cell acts as a bottleneck, strictly disconnecting left and right parts of a given state of the Hilbert space and completely removing any global heat transport. 
Consequently, subspaces labelled by $\{\ho_l\}$ are separated into (i) \emph{conducting channels} where every unit cell is conducting (hence, there exists a continuous path along heat-exchanging units which connect opposite ends of the system), and (ii) \emph{non-conducting states} where at least one unit cell is blocking. 
Within a non-conducting state, blocking cells separate contiguous conducting unit cells, \emph{conducting islands}, where heat transport is possible although non-global. The existence of conducting channels -- and the possibility of heat transport -- in 1D networks~\eqref{eq:2Dhamilt} is controlled by the filling fraction $\delta$. For $1/\nu\leq \delta\leq (\nu-1)/\nu$ conducting and non-conducting channels coexist, otherwise only non-conducting states are present.

%   Figure 1
\begin{figure}%[t!]
    \includegraphics[width=\columnwidth]{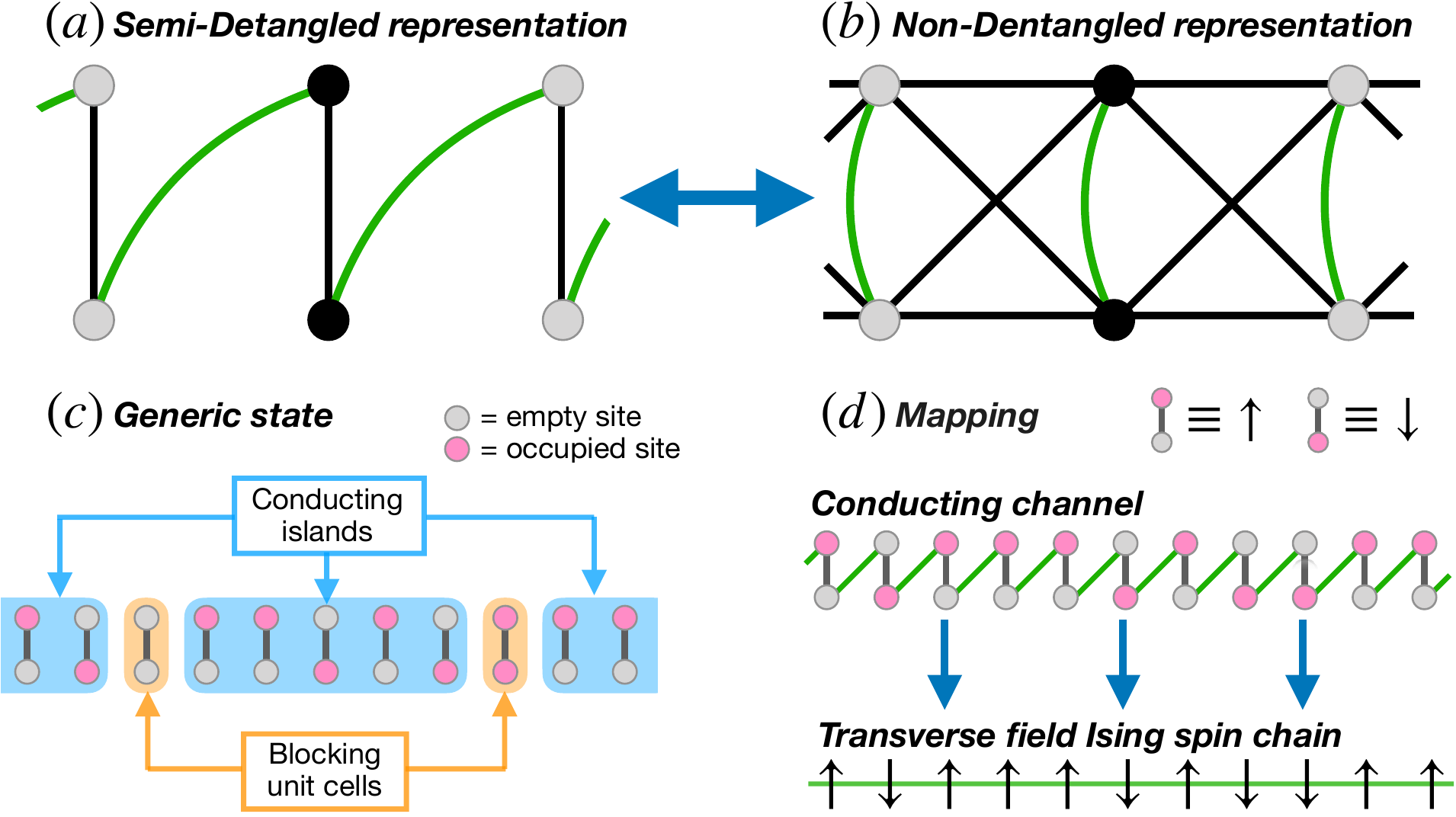}
    \caption{ 
        (a) One dimensional $\nu=2$ MBFBL network Eq.~\eqref{eq:1Dhamilt} with    $\mhsp$ (black lines) and $\mhi$ (green curves). 
        The black circles indicate the unit cell choice.
        (b) Same as (a) with $\mhsp$ non-detangled (cross-stitch). 
        (c) Conducting islands (blue shaded area) and blocking unit cells (orange shaded area) in a generic state of the Hilbert space of $\mh$.
        (d) Visualization of the mapping of $\mh$~\eqref{eq:1Dhamilt} to a transverse field Ising spin chain $\mh^I$~\eqref{eq:1d_ising} in the subspace of conducting channels at $\delta = 0.5$. 
        The green horizontal line indicates the spin-interaction terms.
    }
    \label{fig1}
\end{figure}

The $\nu=2$ case is the minimal testbed setup. In this case, conducting channels are only present at precisely filling fraction $\delta=0.5$ ($\ho_l=1$ in all cells). An example network~\eqref{eq:2Dhamilt} is shown in Fig.~\ref{fig1}(a), the corresponding Hamiltonian $\mh$ reads
\begin{gather}
%    \footnotesize
    \hspace*{-5mm}  
    \mh = \sum_{l}
    \begin{pmatrix}
        \ha_{l}, &\hb_{l} 
    \end{pmatrix}^\dagger
    \begin{pmatrix}
        s & t \\
        t & s
    \end{pmatrix}
    \begin{pmatrix}
        \ha_{l} \\
        \hb_{l}
    \end{pmatrix}
    + V \sum_{l} \hat n_{b,l} \hat n_{a,l+1},
    \label{eq:1Dhamilt}
\end{gather}
where $\ha_{l},\hb_{l}(\ha_{l}^{\dagger},\hb_{l}^{\dagger})$ are the fermionic  annihilation (creation) operators, $\hat n_{a,l}, \hat n_{b,l}$ are the respective particle-number operators and $s,t$ are onsite energies and intracell hopping respectively. Note that this network is related to the (non-detangled) cross-stitch lattice Fig.~\ref{fig1}(b) by local unitary transformations~\cite{danieli2020nonlinear}, that preserve $\mhi$ as density-density interaction~\cite{danieli2020many}. 

A generic non-conducting state is shown in Fig. \ref{fig1}(c), where contiguous conducting islands are separated by blocking unit-cells. Instead, a sample conducting channel is shown in Fig. \ref{fig1}(d).
%Examples of a conducting channel and generic nonconducting channel, where contiguous conducting islands are separated by blocking unit cells, are shown in Fig.~\ref{fig1}(d) and~\ref{fig1}(c) respectively. 
Within the conducting channels
%-- i.e. the subspace defined by $\forall l:~\ho_l =1$ --
that exist at $\delta = 0.5$, the Hamiltonian $\mh$~\eqref{eq:1Dhamilt} maps onto a spin-$\frac{1}{2}$ transverse field Ising model~\cite{stinchcombe1973ising}. 
The mapping is visualized in Fig.~\ref{fig1}(d): we define local spin basis $\ket{\uparrow}\equiv \ha_l\ket{0}$ (one fermion at site $a$) and $\ket{\downarrow}\equiv \hb_l\ket{0}$ (one fermion at site $b$) for each unit cell $l$. In this representation, the onsite terms of $\mh$ become an identity, the hopping terms turn into $\sigma_l^x$, while the interaction reads $(1-\hat\sigma^z_l)(1+\hat\sigma_{l+1}^z)/4$ -- where $\hat\sigma_l^\alpha$ are Pauli matrices. The effective spin-$1/2$ Hamiltonian of the conducting channels reads
\begin{gather}
    \mh_\text{I} = t \sum_l \hat\sigma_l^{x} - \frac{V}{4} \sum_l \hat\sigma_l^{z} \hat\sigma_{l+1}^{z} + \frac{V}{4}
    \left(\hat\sigma_L^{z}-\hat\sigma_1^{z}\right),
    \label{eq:1d_ising}   
\end{gather}
where the last term vanishes for periodic boundary conditions \footnote{In Eq.~\eqref{eq:1d_ising} we neglected the constant energy shift terms which do not impact the dynamics of the system}.  
Note that this mapping can also be done within individual conducting islands of nonconducting channels.
In that case, the boundary term in Eq.~\eqref{eq:1d_ising} depends on whether the blocking cells at the edges of the island are empty or filled -- see~\footnote{For a conducting island with $1\leq l\leq M$ conducting cells, the boundary terms in Eq.~\eqref{eq:1d_ising} read $\frac{1}{4} \big[(\ho_0 - 1)\hat\sigma_1^{z} + (1 - \ho_{M+1})\hat\sigma_{M}^z \big]$, where $\ho_0,\ho_{M+1}$ are the particles number operators in unit cells $l=0,M+1$\label{note1}}.

This mapping of $\mh$~\eqref{eq:1Dhamilt} to the transverse field Ising chain $\mh_\text{I}$~\eqref{eq:1d_ising} implies ballistic heat transport in the conducting channel~\cite{sun2010heat}. 
However, the total dimension of nonconducting channels overpowers the dimension of the conducting channel for diverging number $L$ of unit cells. 
Indeed, the size of the conducting channel is $2^L$ while the full Hilbert space dimension is $\mathcal{D} = {{2L}\choose{L}}\sim 2^{2L}/\sqrt{L}$ for large $L$.
Consequently, the relative dimension $\mathcal{R}$ of the conducting channel with respect to the full Hilbert space $\mathcal{D}$ is $\mathcal{R}\sim\sqrt{L} 2^{-L}$, implying an exponential suppression of heat transport for large $L$. Note that the eigenenergies of the conducting subspace are spread throughout the entire spectrum. 
These results are not specific to the testbed case Eq.~\eqref{eq:1Dhamilt}, but  apply to any choice of $\mhsp$ and $\mhi$ in $\mh$ in Eq.~\eqref{eq:2Dhamilt} that leads to MBFBL.

%   Figure 2
\begin{figure}%[t!]
    \includegraphics[width=\columnwidth]{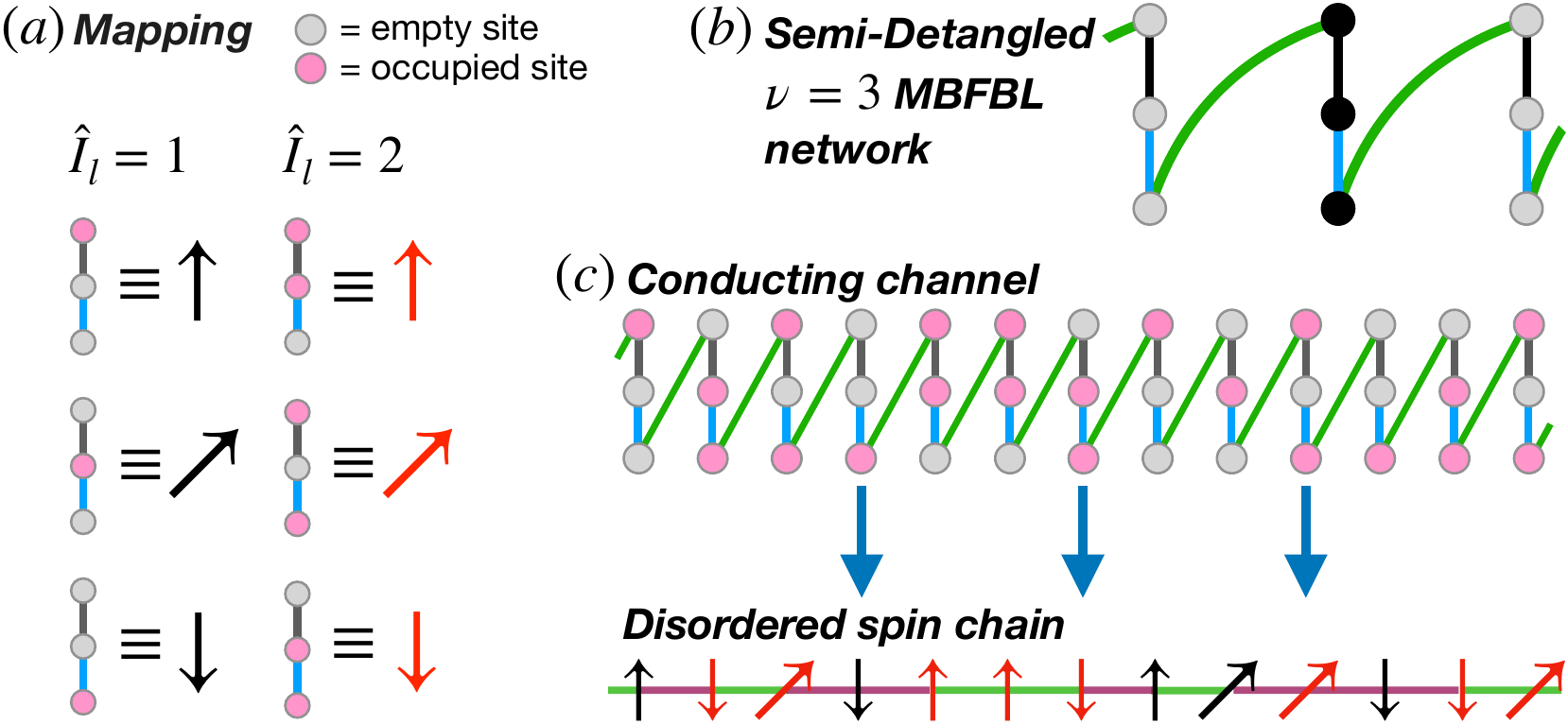}
    \caption{ 
        (a) Mapping of conducting unit cells to spin--1 representation of $\nu=3$ MBFBL networks separated in $\ho_l=1$ and $\ho_l=2$. 
        Black and red colors indicate different spin field components.
        (b) One dimensional sample $\nu=3$ MBFBL network  Eq.~\eqref{eq:2Dhamilt} represented alike Fig.~\ref{fig1}(a). 
        Blue and black lines indicate two different hoppings. 
        (c) Recasting a conducting channel into spin representation via the mapping in (a). 
        Green and magenta horizontal lines represent the spin interaction terms between neighboring unit cells with same or different value of $\ho_l$ respectively.
    }
    \label{fig2}
\end{figure}

In general, for $\nu \geq 3$ conducting channels are present within a range of filling fractions $1/\nu\leq \delta\leq (\nu-1)/\nu$. Similarly to the $\nu=2$ case, the ratio $\mathcal{R}$ between the dimension of the conducting channels and the full Hilbert space dimension decays exponentially in $L$, with the decay rate depending on the number of bands and the filling fraction. For example, for $\nu=3$ the ratio scales as
\begin{gather}
    \mathcal{R}\sim \left\{ \frac{(\delta - 1/3)^{\delta - 1/3}(2/3 - \delta)^{2/3 - \delta}}{\delta^\delta (1 - \delta)^{1 - \delta}} \right\}^{-3L}, 
\end{gather}
where the expression in the parenthesis is strictly greater than one for any  $1/3 \leq \delta \leq 2/3$.
%the base of the exponent is strictly greater than one for any  $1/3 \leq \delta \leq 2/3$. 

For $\nu=3$ systems~\eqref{eq:2Dhamilt}, conducting channels are given by $\ho_l =1,2$ $\forall l$ -- namely, each unit cell contains either one or two fermions.
This corresponds to three degrees of freedom per unit cell, leading to a mapping onto a spin-$1$ model.
%Since one and two fermions on a single unit cell correspond to three degrees of freedom systems, they both can be mapped onto a spin-1 component. 
However, the matrix $H_0$ in~\eqref{eq:2Dhamilt}, that translates into local fields, takes different form for singly or doubly occupied cells~\footnote{The two forms of $H_0$ are equal for specific cases -- \emph{e.g.} with constant diagonal entrees $s$ and equal non-zero off-diagonal entrees $t$}.
%is in general mapped onto two different field components for single- or double-occupied unit cells~\footnote{The two field components are equal for single- or double-occupied unit cells for specific cases of $H_0$ -- {\it e.g.} with constant diagonal entrees $e$ and equal non-zero off-diagonal entrees $t$}. 
Similarly, the mapping of the interaction $\mhi$ depends on the filling fraction of the neighbouring unit cells $\ho_l,\ho_{l+1}$.
%Furthermore, depending on the filling of two neighboring unit cells $\ho_l,\ho_{l+1}$, the interaction $\mhi$ is mapped onto different spin operators~\textbf{IV: potentially write details in SM}.
Consequently, an inhomogeneous distribution of charges in the conducting channel generates an effective disorder in the interaction terms of the spin-$1$ Hamiltonian, and possibly in its local field components (according to $H_0$). 
Such effects also persist for $\nu \geq 4$ networks where in addition to this disorder in local fields and the interactions, different possible fillings of unit cells $1\leq \ho_l\leq \nu-1$ produce spins of different lengths -- yielding an effective disorder in the spin length.
All these effective disorders can hinder and potentially halt transport in  conducting channels: it has been shown that disorder in the interaction can induce MBL~\cite{bar2016many}. 
This mapping
%-- whose local spins are schematically represented in Fig.~\ref{fig2}(a), where black and red colors indicate the different field components -- 
is illustrated in Fig.~\ref{fig2}(b) for a sample MBFBL network, 
%a system which
that extends Eq.~\eqref{eq:1Dhamilt} to $\nu=3$.
The details of the mapping are schematically represented in Fig.~\ref{fig2}(a), where black and red colors indicate the different field components of $H_0$.
The resulting transformation from a sample conducting channel to spin chain is visualized in Fig.~\ref{fig2}(c), where the green and magenta horizontal lines indicate  different interaction operators.

\sect{Higher dimensions}
Unlike the one-dimensional case, a single blocking unit cell does not completely halt global transport in $D \geq 2$. 
Thus, a subspace is a \emph{conducting channel} if there is at least one path formed by conducting cells connecting opposite sites of the network.
Otherwise, a subspace is non-conducting. 
For finite system sizes conducting channels exist only in a limited range of filling fractions $\delta$, which converges to the full available interval $0<\delta< 1$ for $L\rightarrow\infty$ -- \emph{e.g.} for a square lattice of size $L$, conducting channels exist for $1/(\nu L)\leq \delta \leq (\nu L-1)/(\nu L)$.
In the infinite system size limit, let us consider a random state from the particle-number basis.
The probability $p$ for a given unit cell in the state to be conducting (e.g. nor empty $\ho_{\ml} = 0$ nor full $\ho_{\ml} = \nu$) is $p=1 - \delta^\nu - (1 - \delta)^\nu$. 
However because the sampling is performed from the fermionic Hilbert space basis, the events of each of the unit cells being conducting are correlated. 
Nonetheless, these correlations are inversely proportional to the system size and become negligible as $L \rightarrow \infty$ (as we verify  numerically below).
Therefore counting the relative dimension of conducting channels $\mathcal{R}$ with respect to the size of the Hilbert space in a network~\eqref{eq:2Dhamilt} is equivalent to a site-percolation problem, that is covered by the standard Bernoulli percolation theory~\cite{duminil2018sixty}  
%Then, following percolation theory~\cite{saberi2015recent}, 
In a site-percolation problem there exists always a critical value of probability $p_\text{cr}$~\cite{saberi2015recent} which depends on the network geometry, such that $\mathcal{R} \xrightarrow[L \rightarrow \infty]{} 0$ for $p<p_\text{cr}$ (\emph{non-transporting} regime) while $\mathcal{R} \xrightarrow[L \rightarrow \infty]{} 1$ for $p>p_\text{cr}$ (\emph{transporting} regime). 
%Such critical value $p_\text{cr}$ depends on the geometry of the network and its dimensionality. 

In the MBFBL networks~\eqref{eq:2Dhamilt}, the critical transition probability $p_\text{cr}$ depends on the network connectivity $\langle{\ml_1},{\ml_2}\rangle$ that links the detangled unit cells, together with the dimensionality $D$.
Instead, the particular choices of the matrix $H_0$ and interaction coefficients $J_{a,b}^{{\ml_1},{\ml_2}}$ in Eq.~\eqref{eq:2Dhamilt} are irrelevant for $p_\text{cr}$ (so long as the network connectivity is unchanged). 
The critical values of the filling fraction $\delta_\text{cr}$ where a transition between transporting and non-transporting regimes occurs is related to the critical probability $p_\text{cr}$ through
\begin{gather}
    \label{eq:perc_trans}
    1 - \delta_\text{cr}^\nu - (1 - \delta_\text{cr})^\nu = p_\text{cr}.
\end{gather}

%   Figure 3
\begin{figure}%[t!]
    \includegraphics[width=\columnwidth]{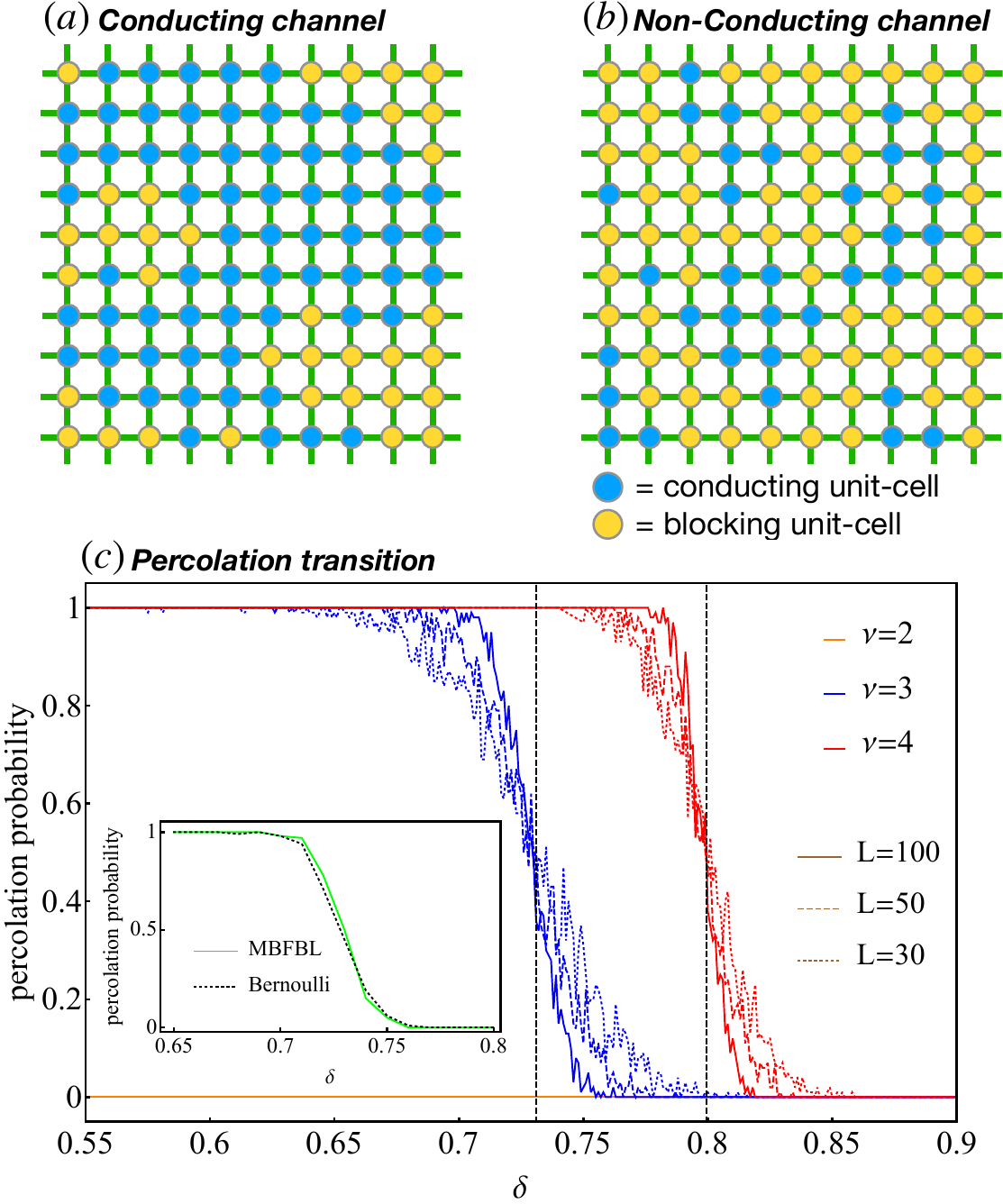}
    \caption{ 
        (a) Typical state from a conducting channel in an equispaced square MBFBL network Eq.~\eqref{eq:2Dhamilt} with FD $\mhi$ (green lines) and SD $\mhsp$ (circles) distinguished in conducting unit cells (blue) and blocked unit cells (orange).
        (b) Same as (a) for a non-conducting subspace. 
        (c) Percolation probability versus filling fraction $\delta$ for different system sizes $L$ and number of sites per unit cell $\nu$ calculated using $100$ Monte-Carlo samples for each point. Vertical dashed lines -- critical filling fraction given by  Eq.~\eqref{eq:perc_trans}. 
        Inset: comparison of Bernoulli and MBFBL percolation models for $\nu=3$, $L=100$. 
    }
    \label{fig3}
\end{figure}

We start with the simplest case and consider a $D=2$ Hamiltonian $\mh$~\eqref{eq:2Dhamilt} arranged as a square lattice.
In Fig.~\ref{fig3} we schematically illustrate conducting channels (a) and states from non-conducting subspace (b).
The site-percolation critical value in this network is $p_\text{cr} \sim 0.59$~\cite{jacobsen2015critical}. 
In Fig.~\ref{fig3}(c) we show the numerically computed percolation probability as a function of the filling fraction $\delta$ for several system sizes $L$ and number of sites per unit cell $\nu$.
The inset of Fig.~\ref{fig3}(c) compares the percolation probabilities calculated from the Bernoulli percolation model and from the direct sampling of the fermionic Hilbert space. 
The results are in excellent agreement justifying the application of the Bernoulli model. 
For $\nu=2$ the probability that a unit cell is conducting is $p\leq 0.5$ for any $\delta$, with the maximum $p=0.5$ being reached at $\delta=0.5$ -- hence, the transition value $p_\text{cr}\sim 0.59$ is never reached, the network never undergoes percolation transition and global transport is suppressed for all filling fractions $\delta$. 
Numerical simulations confirm this conclusion for $\nu=2$.
%the absence of transition for any filling fraction $\delta$. 
For $\nu=3,4$ numerical simulations give clear evidences of a percolation transition upon varying $\delta$, again in excellent agreement with the critical values predicted by Eq.~\eqref{eq:perc_trans}.

In $D=2$ the majority of lattices -- kagome, honeycomb, octagon -- have critical values $p_\text{cr} \gneqq 0.5$~\cite{suding1999site}.
Exceptions include networks with further than nearest-neighbor terms~\cite{majewski2006square}, chimera~\cite{melchert2016site}, and triangular lattice~\cite{smirnov2001critical}. 
Hence, in two dimensions, most of $\nu=2$ MBFBL networks $\mh$ avoid percolation transition (and therefore any transport is absent) since the maximum probability of a unit cell to be conducting is $p=0.5$.
As $D$ increases, the critical transition $p_\text{cr}$ in a given class of lattices typically decreases: for example $D\geq 3$ hypercubic networks $\mh$ percolation transition occurs for any $\nu\geq 2$~\cite{saberi2015recent}.
However, exotic networks on which $\mh$ avoid percolation transition exist: for instance, $\nu=2$ MBFBL networks on 3D cubic-oxide or 3D silicon-dioxide~\cite{tran2013percolation,yoo2014site}.

Geometric percolation theory yields regions of filling fraction where the number of states from conducting channels is dominating over those from the non-conducting ones. 
However, quantum transport in these regions can still be hindered due to several additional factors, that become clear once mappings to spin models, that generalize the $D=1$ case are considered. These factors follow from the local constraints $\{\ho_{\ml}\}$. 
Firstly, as discussed for the spin mapping within the conducting subspace in $D=1$, distributions of particles in the conducting cells can generate disorder in the interactions, the local spin fields (for $\nu \geq 3$), and even the spin lengths (for $\nu \geq 4$). 
Secondly, percolating clusters in conducting channels have in general a highly irregular fractal structure~\cite{hermann1984building}, yielding edge scattering. 
This source of scattering -- already present in classical percolation -- is in our context further amplified by the inhomogeneous arrangements of blocking cells in the network. 
This yields additional, effectively random fields in the spin model in the conducting cells at the edges of the percolating cluster. 
Finally, let us point out that the existence of dead-end bonds which fully reflect waves and lead to an effective increase of the percolation threshold with respect to the classical one -- a phenomenon called \textit{quantum percolation}~\cite{meir1991quantum, schubert2009quantum} -- has been demonstrated for single-particle quantum percolation clusters.
Thus, the classical values of critical probability $p_\text{cr}$ discussed above could only be a lower bound on the effective quantum threshold values for the MBFBL networks.

\sect{Conclusions and Perspectives} 
In this work we studied transport features of translationally invariant interacting many-body flatband (MBFBL) networks~\cite{danieli2020many,kuno2020flat,orito2020exact}. 
While charges are strictly confined in these models due to local constraints, heat is in principle allowed to flow along conducting channels -- leading to site-percolation transitions in the networks. 
We found that in 1D the number of conducting channels either vanishes or decays exponentially with the system size, hence suppressing any  transport. 
In higher dimensions, the number of conducting channels can undergo a percolation transition upon tuning the filling fraction $\delta$, potentially triggering heat transport. 
However, the presence of the transition depends on the network type, the number of single-particle energy bands $\nu$, and the dimensionality $D$. 
While we explicitly verified this conclusions for the case of spinless fermions, these results hold for any type of many-body statistics.

Our analysis -- rooted in geometric percolation theory, alongside recent findings of localization/delocalization transition in 2D quantum link models~\cite{karpov2020disorderfree} -- overcomes the complexity typical of $D\geq 2$ many-body systems which render their computational analysis impractical nowadays.   
Importantly, we highlight classes of disorder-free many-body systems in two and higher spatial dimensions where any type of transport is completely suppressed for any filling fraction $\delta$. 
Even in those networks that feature conducting phases, quantum transport is hurdled and potentially halted by effective disorder and edge effects resulting from the local constraints. 
Exploring these quantum effects and investigating the distinction between classical and quantum percolation transitions~\cite{meir1991quantum, schubert2009quantum} are crucial future developments, which may have important repercussions for the research in the field of quantum transport. 
These results and open quests emphasize MBFBL networks as experimentally realizable~\cite{fang2012realizing,mukherjee2018experimental,gladchenko2009superconducting} valid platforms to explore novel phenomena in quantum systems -- highlighting jointly with Refs.~\cite{tovmasyan2018preformed,tilleke2020nearest,danieli2020quantum,hart2020compact,mcclarty2020disorder,kuno2020flat_qs,daumann2020manybody,khare2020localized,danieli2021manybody,danieli2020many,kuno2020flat,orito2020exact} the progressively growing relevance of flatbands in the realm of many-body physics. 

\sect{Acknowledgments}
This work was supported by the Institute for Basic Science (Project number IBS-R024-D1). We thank Ivan Khaymovich for helpful discussions. 

\bibliography{flatband,mbl,general}

\end{document}